\documentclass[12pt]{article}
\columnwidth\textwidth

\begin{document}

\newcommand{\be}{\begin{equation}}
\newcommand{\ee}{\end{equation}}
\newcommand{\beann}{\begin{eqnarray*}}
\newcommand{\eeann}{\end{eqnarray*}}
\newcommand{\bea}{\begin{eqnarray}}
\newcommand{\eea}{\end{eqnarray}}
\newcommand{\nn}{\nonumber}
\newcommand{\ben}{\begin{enumerate}}
\newcommand{\een}{\end{enumerate}}
\newtheorem{df}{Definition}
\newtheorem{thm}{Theorem}
\newtheorem{lem}{Lemma}
\newtheorem{prop}{Proposition}
\begin{titlepage}

\noindent
\hspace*{11cm} BUTP-97/9 \\
\vspace*{1cm}
\begin{center}
{\LARGE Gauge Invariant Hamiltonian Formalism for Spherically
Symmetric Gravitating Shells}

\vspace{0.5cm}

P. H\'{a}j\'{\i}\v{c}ek \\
Institute for Theoretical Physics \\
University of Bern, Sidlerstrasse 5, CH-3012 Bern, Switzerland \\
\vspace*{0.5cm}

J. Bi\v{c}\'{a}k \\
Department of Theoretical Physics \\
Faculty of Mathematics and Physics, Charles University \\
V Hole\v{s}ovi\v{c}k\'{a}ch 2, 18000 Prague, Czech Republic \\
\vspace*{0.5cm}

September 1997 \\ \vspace*{0.5cm}

\nopagebreak[4]

\begin{abstract}
  The dynamics of a spherically symmetric thin shell with arbitrary rest mass
  and surface tension interacting with a central black hole is studied. A
  careful investigation of all classical solutions reveals that the value of
  the radius of the shell and of the radial velocity as an initial datum does
  not determine the motion of the shell; another configuration space must,
  therefore, be found. A different problem is that the shell Hamiltonians used
  in literature are complicated functions of momenta (non-local) and they are
  gauge dependent. To solve these problems, the existence is proved of a gauge
  invariant super-Hamiltonian that is quadratic in momenta and that generates
  the shell equations of motion. The true Hamiltonians are shown to follow
  from the super-Hamiltonian by a reduction procedure including a choice of
  gauge and solution of constraint; one important step in the proof is a lemma
  stating that the true Hamiltonians are uniquely determined (up to a
  canonical transformation) by the equations of motion of the shell, the value
  of the total energy of the system, and the choice of time coordinate along
  the shell. As an example, the Kraus-Wilczek Hamiltonian is rederived from
  the super-Hamiltonian. The super-Hamiltonian coincides with that of a
  fictitious particle moving in a fixed two-dimensional Kruskal spacetime
  under the influence of two effective potentials. The pair consisting of a
  point of this spacetime and a unit timelike vector at the point, considered
  as an initial datum, determines a unique motion of the shell.

\end{abstract}

\end{center}

\end{titlepage}

\section{Introduction}
\label{sec:intro}
One of the most spectacular and at the same time the most mysterious
phenomena in nature is the gravitational collapse. It produces the
largest fireworks in the sky and it ends up in a singularity, to which
no known physical laws seem to be applicable. The enigma is enhanced
by the appearance of horizons before the final stages of 
collapse. The horizon has some regularizing effect; for example, it
prevents the divergence of the total energy due to concentration of
matter (unlike the electrodynamics). However, the horizon leaks and
there is another mystery: does the Hawking effect lead to violation of
unitarity or not?

The theory of gravitational collapse is difficult, not least because the
phenomenon seems to belong to the high energy regime of quantum
gravity and the progress in quantum gravity is slow. Thus,
impatient people try to capture the core of the problem by working
with simplified models. This is also the line of the present paper.

Let us consider the simplest models of the collapse: spherically
symmetric thin shells (in general with a surface tension) falling in a
field of a spherical black hole, and Oppenheimer-Snyder stars. Let us
restrict the dynamical systems to their minimum: a fixed ``amount of
matter'' without any internal degrees of freedom.  For example, the
total rest mass of dust shells and of the Oppenheimer-Snyder star is
fixed. Let us, finally, suppress all fields different from
gravity. The resulting dynamical systems have just one degree of
freedom; this can, at least locally, be described by the radius $r$ of
the matter part of the system (of the shell, of the Oppenheimer-Snyder
star).

Such or similar simple systems have been often utilized in
literature. For instance, in the treatments investigating 
relativistic dynamics of spherical domain walls (sometimes called
bubbles) \cite{bubbles}, models of gravitational collapse
\cite{PH-coll}, back reaction in the Hawking effect \cite{K-W}, or
models of quantum black holes \cite{M-Y}, \cite{berezin}, 
\cite{ital}.

If we study these papers, we find rather surprisingly that the theory
of even such simple systems is afflicted with severe problems. Thus,
most papers start with guessing the Hamiltonian (or super-Hamiltonian)
of the system to be studied from its equations of motion (eg.\
\cite{FGG}, \cite{berezin}, \cite{PH-coll}). However, it is well-known
that a Hamiltonian and, therefore, an action is not uniquely
determined by the equations of motion that it is to generate (this is
the so-called inverse problem of variational calculus). In some
papers, the Hamiltonians are derived from the first principles but the
derivation is extremely complicated \cite{K-W}. The results of both
guesses and derivations are, as a rule, very complicated, non-local
Hamiltonians. A complicated Hamiltonian {\em is} a problem, if we are
going to base a quantization on it and to calculate anything of
interest. Sometimes, variables are chosen such that the corresponding
canonical formalism becomes manageable but this simplicity is often
bought by disregarding large portions of the space of solutions (the
physical phase space); this seems to be the case in
\cite{ital}. Finally, and this is really bad: different choices of
time coordinate along the matter boundary lead to very different forms
of Hamiltonian for the same system, thus making the corresponding
naive quantizations non-equivalent, and so gauge dependent.

The present paper is an attempt to deal with all these problems in the case of
the shells. First, we show that a Hamiltonian of a system with one degree of
freedom is uniquely determined (up to a canonical transformation) if its value
is prescribed to be the total energy of the system. Very simple methods to
calculate such Hamiltonians are presented. Second, we present a careful study
of all classical solutions of each respective system. This study reveals that
the systems possess more states (ie.\ classical solutions) and admit more
symmetry than various quantum approaches usually assume. There are not only
bound states, but also scattering ones, and there is a number of states beyond
horizon. The symmetries are related to isometries of Kruskal spacetime. This
implies that the motion of the system is not (globally) determined by the
value of the pair $(r,\dot{r})$, and that the energy of the system is not a
(globally) well-defined function of $r$ and $\dot{r}$. Another result is a
one-to-one correspondence between the set of classical solutions and that of
certain trajectories on a certain spacetime.\footnote{S.~K.~Blau,
  E.~I.~Guendelman and A.~H.~Guth \cite{BGG}, following the suggestion of
  S.~Coleman, discussed the classical solution of the equation of motion of
  the spherical domain wall as the solution of the equation of motion of a
  particle moving in one dimension under the influence of a potential.}  The
spacetime is closely associated with the classical solution; for the shell in
the field of a black hole, it is the maximal extension of the $uv$-surface of
the the part of Kruskal spacetime inside the shell and the trajectories
coincide with those of the shell in this spacetime. The dynamics of the
trajectories can be obtained from a super-Hamiltonian on which three
conditions are imposed: it should be gauge invariant, at most quadratic in
momenta and it should reproduce all symmetries in the space of solutions.
Under these conditions, the super-Hamiltonian is unique (up to an overal 
factor). In this way, the variables describing the
gravitational field---which are pure gauges in the spherically symmetric
case---are reduced away without any choice of coordinate conditions. A similar
super-Hamiltonian has been obtained and used for the quantization of thin dust
shells with the flat spacetime inside \cite{H-K-K}. It follows from the
uniqueness of the true Hamiltonians that they can be obtained from our
super-Hamiltonian by reduction procedures.

The plan of the paper is as follows. In Sec.\ \ref{sec:kruskal}, we
briefly describe the relevant properties of Kruskal spacetime, in
particular the symmetries. In Sec.\ \ref{sec:hamilt}, we prove the
uniqueness of Hamiltonian discussed above, present some simple methods
to derive such Hamiltonians and show two examples of complicated and
gauge-dependent Hamiltonians. 
Sec.\ \ref{sec:shell} is devoted to massive
thin shells with arbitrary surface tension. By including tension, we
can describe shells made from fundamental fields such as scalar fields
in the bubbles arising in cosmological phase transitions
\cite{bubbles}. Sec.\ \ref{sec:null} collects what we need about null
shells. In Sec.\ \ref{sec:particle}, we write down a Hamiltonian
formalism for the shells. In Sec.\ \ref{sec:out}, starting from our
super-Hamiltonian, we derive the Hamiltonians that describe the motion
of null shells in the Kruskal spacetime external to the shell (like
Ref.\ \cite{K-W}).

\section{Kruskal spacetime}
\label{sec:kruskal}
All generic (not extreme) black hole spacetimes contain a region which
has the same topology as the Kruskal spacetime. It includes two
identical stationary asymptotically flat submanifolds that are
causally separated by two crossing horizons, and two identical
dynamical submanifolds that are time inversions of each other. For the
astrophysical description of a stellar collapse, only a half of this
spacetime is necessary and relevant. What can be the consequence of
the existence of the other half?  Apart from the discussions of
primordial wormholes or white holes, it seems that it may have some
importance in the quantum theory: indeed, for example the popular
explanation of Hawking effect uses the states beyond the horizon
because they have negative energy with respect to infinity. Let us
start by briefly describing the Kruskal spacetime, restricting
ourselves to the two-dimensional version of it.

Consider ${\mathbf R}^2$ with double null coordinates $u$ and $v$ and the
metric
\be
  ds^2 = \frac{(4\mbox{G}E)^2}{F}\mbox{e}^{-F}(-du\,dv),
\label{metric}
\ee
where $E > 0$, G is the Newton constant and $F=F(-uv)$ is
defined by its inverse,
\[
  F^{-1}(x) := (x - 1)\mbox{e}^x
\]
in the interval $x \in (0,\infty)$ and $-uv \in (-1,\infty)$.
The metric is well-defined and analytic in the region $-uv \in
(-1,\infty)$. This part of ${\mathbf R}^2$ together with the metric
(\ref{metric}) is called Kruskal spacetime $({\mathcal M},g)$. 

The Kruskal spacetime is time and space orientable. We will call {\em
future} the orientation in which the function $(u+v)$ is increasing,
and {\em right} in which $(v-u)$ is increasing. There are two
independent discrete isometries that invert these orientations: T,
defined by $u \mapsto -v$, $v \mapsto -u$ and C by $u \mapsto v$, $v
\mapsto u$. Their composition I = CT = TC is called the {\em Kruskal
inversion}; I inverts both orientations. Observe that C is a new
transformation, enabled by the topology of the Kruskal manifold. C has
nothing to do with parity; indeed, the parity transformation P leaves
both $u$ and $v$ unchanged but transforms $\vartheta$ and $\varphi$ by
$\vartheta \mapsto \pi/2 - \vartheta$ and $\varphi \mapsto \varphi +
\pi$ in the full four-dimensional manifold.

There is also a continuous one-dimensional group of isometries
generated by the Killing vector field
\be
  \xi := \frac{1}{4\mbox{G}E} \left(-u\frac{\partial}{\partial u} +
  v\frac{\partial}{\partial v} \right).
\label{killing}
\ee
If $g(s)$ is an arbitrary element of this group, then
\[
  \mbox{T}g(s)\mbox{T} = g(-s),\quad \mbox{C}g(s)\mbox{C} =
  g(-s),\quad \mbox{I}g(s)\mbox{I} = g(s).
\]
The signature and orientation of $\xi$ changes over $M$. Let us define
four quadrants in $M$ as follows:
\beann
  Q_I & := & \{(u,v)\,|\,u < 0, v > 0\}, \\
  Q_{II} & := & \{(u,v)\,|\,u > 0, v < 0\}, \\
  Q_{III} & := & \{(u,v)\,|\,u > 0, v > 0\}, \\
  Q_{IV} & := & \{(u,v)\,|\,u < 0, v < 0\}.
\eeann
Then, the orientation and signature of $\xi$ is given by the table
\begin{center}
\begin{tabular}{l||l|l}
    & signature & orientation \\ \hline \hline
$Q_I$ & timelike & future \\ \hline
$Q_{II}$ & timelike & past \\ \hline
$Q_{III}$ & spacelike & right \\ \hline
$Q_{IV}$ & spacelike & left \\
\end{tabular}
\end{center}
The group generated by T, C, and $\xi$ is called the {\em Kruskal group},
$G_K$. 

Two important functions, $t$ and $r$, on $M$ are defined as follows:
\[
  r := 2\mbox{G}E\,F(-uv)
\]
everywhere,
\[
  t := 4\mbox{G}E\,\mbox{arctanh}\frac{u+v}{-u+v}
\]
in $Q_I \cup Q_{II}$, and
\[
  t := 4\mbox{G}E\,\mbox{arctanh}\frac{-u+v}{u+v}
\]
in $Q_{III} \cup Q_{IV}$. The pair $(t,r)$ of functions can be
considered as coordinates within each of the four quandrants; these
are usual Schwarzschild coordinates. We have the following
transformations:
\[
  dr = -2\mbox{G}E\frac{\mbox{e}^{-F}}{F}(vdu + udv)
\]
on $M$, and
\[
  dt = -2\mbox{G}E\,\frac{vdu - udv}{uv}
\]
on $Q_I \cup Q_{II}$, and
\[
  dt = 2\mbox{G}E\,\frac{vdu - udv}{uv}
\]
on $Q_{III} \cup Q_{IV}$, where we can also set
\[
  F(-uv) = \frac{r}{2\mbox{G}E}, \quad -uv = \frac{r -
  2\mbox{G}E}{2\mbox{G}E}\,\mbox{e}^{\frac{r}{2\mbox{\tiny{G}}E}}.
\]
Within the quadrants and in the Schwarzschild coordinates, we have $\xi =
\partial/\partial t$.

The behaviour of the functions $t$ and $r$ with respect to the Kruskal
group is as follows. Let $p \in M$ and $g(s)$ be an element of the
component of identity of $G_K$. Then
\beann
  (t(g(s)p),r(g(s)p)) & = & (t(p) + s,r(p)), \\
  (t(\mbox{T}p),r(\mbox{T}p)) & = & (- t(p),r(p)), \\
  (t(\mbox{C}p),r(\mbox{C}p)) & = & (-t(p),r(p)), \\
  (t(\mbox{I}p),r(\mbox{I}p)) & = & (t(p),r(p)).
\eeann
Thus, $r$ is an invariant function of $G_K$, and any invariant
function of $G_K$ must be a function of $r$. Moreover, on the part of
$M$ covered by the four quadrants, the fixed value $(t,r)$ of the pair
of functions $t$ and $r$ defines exactly two points, $p$ and I$(p)$.

\section{Hamiltonians of simple systems}
\label{sec:hamilt}
In this section, we consider spherically symmetric systems that a) are
sufficiently simple so that they have just one degree of freedom and
b) the matter has a well defined external boundary so that the
geometry outside this boundary coincides with a portion of the
Kruskal spacetime. We study the ways, in which a Hamiltonian
generating the motion of such systems can be obtained from first
principles (that is, from the Einstein-Hilbert action to which some
matter action is coupled). Our aim is to show that such Hamiltonians
are a) unique in certain sense if the time coordinate is chosen, b)
very complicated, and c) gauge dependent.

In order to obtain a true Hamiltonian, we have to reduce the
system. This, roughly, is the following procedure. First, we have to
adapt the coordinates to the spherical symmetry. Second, we have 
to choose a foliation given by levels of a function $T$ on the phase
space and a radial coordinate $X$ along the folii; then, we 
have to solve the constraints for the momenta $p_T$ and $p_X$
conjugate to $T$ and $X$. The result can be written as
\[
  p_T + H(q,p,T) = 0,
\]
where $q$ and $p$ are the coordinates on the physical phase space of
the system; the true Hamiltonian is $H(q,p,T)$ (see, eg.\ \cite{ADM}).

Since the spacetime outside the matter has the Kruskal geometry, we
can choose the foliation such that it assymptotically coincides with
the Schwarzschild foliation. It is well-known (\cite{ADM}, \cite{RT})
that the value of the true Hamiltonian then coincides with the ADM
energy $E$ ($E = \mbox{G}M$, where $M$ is the Schwarzschild mass
parameter of the external Kruskal spacetime). Observe that the
foliation can be different from the foliation by the Schwarzschild
time coordinate at the boundary of the matter. Thus, the foliation
will determine a time parametrization of the boundary. Let us denote
the time parameter by $s$; the value $s(p)$ of the parameter at a
given point $p$ of the boundary coincides with the value $T(p)$ of the
function $T$ at $p$. In fact, $s$ can be an arbitrary time parameter
along the boundary.  In this way, each motion of the system determines
a function $r(s)$, where $r$ is the Schwarzschild radius of the
boundary at the value $s$ of the time parameter. As the ADM energy is
always conserved, $H$ cannot depend explicitly on $T$.

Since the system has just one degree of freedom, the function $r(s)$
can contain all information about the motion (at least locally). In
this case, we can choose $r$ and its conjugate momentum $p$ as the
canonical coordinates on an open set in the physical phase space of
the system. The reduction procedure will then result in the true
Hamiltonian of the form $H(r,p)$.

On the other hand, one can also study the dynamical equations of the
system. They follow from Einstein's equations and from the
stress-energy tensor conservation for the matter. Solving the gravity
part of these equations by the Schwarzschild metric, a non-trivial
radial equation remains:
\be
  \dot{r} = \rho(r,E),
\label{em-rho}
\ee
where the dot denotes the derivative with respect to the parameter $s$.
The system can possess different sectors, so $\rho$ can also depend on
a few discrete parameters. We stress that the function $\rho$ is
completely determined by the dynamical equations; to derive its form,
one does not need to employ the canonical formalism and its reductions.

The crucial question now is: How many Hamiltonians of the form $H(r,p)$
will lead to the equation of motion (\ref{em-rho})? It is well-known
that there are many such Hamiltonians \cite{unique-H}. However, if we
impose the condition on $H$ that its value coincides with the total
energy $E$, then the possible Hamiltonians are strongly
limited. Indeed, suppose that $H(r,p)$ is such a Hamiltonian. Then, we
must have
\[
  \dot{r} = \frac{\partial H}{\partial p}(r,p).
\]
This equation represents the transformation between the variables
$(r,\dot{r})$ and $(r,p)$. Thus, the Eq.\ (\ref{em-rho}) implies that
\[
  \frac{\partial H}{\partial p} = \rho(r,H).
\]
This can be considered as a differential equation for $H$; it is even
a separable first-order ordinary differential equation as $r$ 
plays the role of a parameter:
\be
  dp = \frac{dH}{\rho(r,H)}.
\label{separ}
\ee
Hence, the general solution has the form:
\be
  p = p_0(r) + \int\frac{dH}{\rho(r,H)},
\label{moment}
\ee
where $p_0(r)$ is an arbitrary function of $r$. Solving Eq.\
(\ref{moment}) for $H$, we obtain a family of Hamiltonians of the
form:
\[
  H = H(r,p-p_0(r)).
\]
The result is clearly non-unique, but this non-uniqueness, represented
by the function $p_0(r)$, can be regarded as the symmetry of the
formalism under the transformation
\be
  r \mapsto r,\quad p \mapsto p - p_0(r).
\label{can-transf}
\ee
It is clear that we always have the freedom to perform such a
canonical transformation in the classical Hamiltonian formalism (to
simplify the equations, for example). More important, however, is the
fact that the transformation (\ref{can-transf}) belongs to the small
class of canonical transformations that can be implemented by a
unitary transformation in the quantum theory. Indeed, suppose that we
represent the Heisenberg commutation relations for $r$ and $p$ as
usual:
\[
  (\hat{r}_1\psi_1)(r) = r\psi_1(r),\quad (\hat{p}_1\psi_1)(r) =
  -i\psi'_1(r), 
\]
and that the wave functions $\psi_1(r)$, $\phi_1(r)$, etc.\ span the
Hilbert space with the scalar product
\[
  (\psi_1,\phi_1) = \int_0^\infty dr\,\psi^*_1(r)\phi_1(r).
\]
Next, we consider another copy of the same Hilbert space, but denote its
elements by $\psi_2(r)$, $\phi_2(r)$, etc.\ and define a unitary
transformation between the spaces by
\[
  \psi_2(r) = \mbox{e}^{-iY(r)}\psi_1(r),
\]
where $Y(r)$ is a fixed differentiable function. The form of the
operators $\hat{r}_2$ and $\hat{p}_2$ is given by
\[
  \hat{r}_2 = \mbox{e}^{iY(r)}\hat{r}_1\mbox{e}^{-iY(r)} = \hat{r}_1,
\]
\[
  \hat{p}_2 = \mbox{e}^{iY(r)}\hat{p}_1\mbox{e}^{-iY(r)} = \hat{p}_1 -
  Y'.
\]
Hence, the choice $Y(r) = \int dr\,p_0(r)$ shows our claim. In this
sense, the Hamiltonian $H$ can be considered as uniquely determined by
1) the equations of motion, 2) the value of energy and 3) the gauge.

The description of dynamics based on a true Hamiltonian has,
however, the two disadvantages discussed in the Introduction:
complicated form and gauge dependence.

Let us consider a null dust shell (see Sec.\ \ref{sec:null}); it always
moves along a null geodesic in both Kruskal space\-times, the internal
as well as the external one with respect to the shell. In the
Schwarz\-schild co\-ordinates, Eq.\ (\ref{Smetric}) holds, and so the
functions $t_+(s)$ and $r(s)$ that describe the trajectory of the
shell in the external Kruskal spacetime must satisfy the equation
\[
  -f_+(r)\dot{t_+}^2 + f_+^{-1}(r)\dot{r}^2 = 0.
\]
Hence,
\[
  \left(\frac{dr}{dt_+}\right)^2 = \left( 1 -
  \frac{2\mbox{G}E}{r}\right)^2,
\]
and
\[
  \rho(r,E) = \epsilon\left|1 - \frac{2\mbox{G}E}{r}\right|,
\]
where $\epsilon = \pm 1$ distinguishes the ingoing from the outgoing
shells. In this case, Eq.\ (\ref{moment}) becomes 
\[
  p = p_0 - \epsilon\frac{r}{2\mbox{G}}\ln\left|1 -
  \frac{2\mbox{G}E}{r}\right|, 
\]
and its solution with respect to $E$ yields the form of the Hamiltonian: 
\be
  H = \frac{r}{2\mbox{G}}\left\{1 -
  \exp\left[-\epsilon\frac{2\mbox{G}}{r}(p - p_0(r))\right]\right\}.
\label{Hschw}
\ee
This {\em is} a very complicated, non-local Hamiltonian. 

Next, we choose the Kraus-Wilczek coordinates $\tau$ and $r$ on the
external Kruskal spacetime which are defined by \cite{K-W} 
\be 
  \tau = t_+ +
  2\sqrt{2\mbox{G}Er} + 2\mbox{G}E\ln\left|\frac{\sqrt{r} -
  \sqrt{2\mbox{G}E}}{\sqrt{r} + \sqrt{2\mbox{G}E}}\right|,
\label{KW-t}
\ee
so that the metric takes the spatially flat form
\[
  ds^2  = - f_+ d\tau^2 + 2\sqrt{\frac{2\mbox{G}E}{r}}drd\tau + dr^2
  + r^2d\Omega^2.
\]
The equation of a radial null geodesic in this metric reads
\[
  \left(\frac{dr}{d\tau}\right)^2 + 2\sqrt{\frac{2\mbox{G}E}{r}}
  \frac{dr}{d\tau} - 1 + \frac{2\mbox{G}E}{r} = 0
\]
and its solution is
\[
  \frac{dr}{d\tau} = \epsilon - \sqrt{\frac{2\mbox{G}E}{r}},
\]
where $\epsilon$ has the same meaning as above. The Eq.\
(\ref{moment}) now reads
\be
  \mbox{G}p = \mbox{G}p_0(r) - \sqrt{2\mbox{G}Er} -
  \epsilon r\ln \left|\frac{\sqrt{2\mbox{G}E} - \epsilon
  \sqrt{r}}{\sqrt{r}}\right|.
\label{mom-KW}
\ee
This will coincide with the Kraus-Wilczek momentum $p_c$ if one matches
the units and the notation and if $p_0(r)$
is chosen properly:
\[
  \mbox{G}p_0(r) = \sqrt{2\mbox{G}E_-r} +
  \epsilon r\ln \left|\frac{\sqrt{2\mbox{G}E_-} - \epsilon
  \sqrt{r}}{\sqrt{r}}\right|,
\]
where $E_-$ is the Schwarzschild energy of the Kruskal spacetime
inside the shell. We observe again that the Hamiltonian is very
complicated (there is no ecxplicit formula this time that would express
the solution of the Eq.\ (\ref{mom-KW}) with respect to $E$), and it
is also very different from Eq.\ (\ref{Hschw}).

In order to solve this problem, we shall try to find a {\em
super-Hamiltonian}, which will be gauge independent, at most quadratic
in momenta, and such that its reduction based on a suitable gauge will
give each of the complicated Hamiltonians. Such a super-Hamiltonian
can either be itself directly used for quantization, or define the
factor ordering of the complicated Hamiltonians.

\section{Massive shells}
\label{sec:shell}
In this section, a careful study of the states of massive spherically
symmetric dust shells (in general with arbitrary surface tension)
will be given. In the following section, the null dust shells will emerge as a
(degenerate) simple limit case.

\subsection{The trajectory of the shell}
\label{sec:normal}
Consider two Kruskal manifolds ${\mathcal M}_1$ and ${\mathcal M}_2$
with energies $E_-$ and $E_+$ containing isometric timelike
hyper-surfaces $\Sigma_1$ and $\Sigma_2$ which divide each ${\mathcal
M}_i$ in two parts, left ${\mathcal M}_{i-}$, and right ${\mathcal
M}_{i+}$. Let us solder ${\mathcal M}_{1-}$ to ${\mathcal M}_{2+}$
along $\Sigma_1$ and $\Sigma_2$. Most points in the resulting surface
$\Sigma := \Sigma_1 \cap \Sigma_2$ in the resulting spacetime will
have a neighbourhood in which we can introduce Schwarzschild
coordinates on both sides. The metric left ($-$) and the metric right
($+$) is given by ($\epsilon = \pm 1$)
\be
  ds_\epsilon^2 = -f_\epsilon(r)dt_\epsilon^2 + f_\epsilon^{-1}(r)dr^2 +
  r^2(d\theta^2 + \sin^2\theta\,d\phi^2),
\label{Smetric}
\ee
where
\[
  f_\epsilon := 1 - \frac{2\mbox{G}E_\epsilon}{r};
\]
we assume that $r$, $\theta$ and $\phi$ are continuous functions
through $\Sigma$, however, in general, $t_+ \neq t_-$ at $\Sigma$. The
surface $\Sigma$ is given by
\be
  t_\epsilon = t_\epsilon(s),\quad r = r(s),\quad \theta =
  \vartheta,\quad \phi = \varphi,
\label{Sigma}
\ee
where $s$, $\vartheta$ and $\varphi$ are coordinates on $\Sigma$ and
$s$ is the proper time along the radial generators of $\Sigma$. Let us denote
by 
\[
T_\epsilon^\mu := 
(\dot{t}_\epsilon, \dot{r}, 0, 0)
\] 
the tangent vector along these generators. Since $s$ is the proper time,
we obtain 
\be
  f_\epsilon \dot{t}_\epsilon = \tau_\epsilon\sqrt{f_\epsilon +
  \dot{r}^2}, 
\label{e}
\ee 
where $\tau_\epsilon :=
\mbox{sign}(f_\epsilon\dot{t}_\epsilon)$. The meaning of the left hand
side is $ f_\epsilon\dot{t}_\epsilon =
-g^\epsilon_{\mu\nu}T^\mu_\epsilon\xi^\nu_\epsilon$, where
$\xi^\mu_\epsilon$ is the Killing vector (\ref{killing}).  This would
be a total energy with respect to the right infinity for a particle
moving in the respective Kruskal spacetime, and it would be a
conserved quantity for a geodesic motion.  Let $m_\epsilon^\mu$ be the
normal vector to $\Sigma$, $m_+$ being external to ${\mathcal M}_{2+}$,
and $m_-$ to ${\mathcal M}_{1-}$; we assume $m_\epsilon^\mu$ to be
normalized, or
\[
  f_\epsilon\dot{t}_\epsilon m^0_\epsilon -
  f^{-1}_\epsilon\dot{r}m^1_\epsilon  = 0,
\]
\[
  f_\epsilon (m^0_\epsilon)^2 -
  f^{-1}_\epsilon (m^1_\epsilon)^2  = -1.
\]
It follows that
\be
  m_\epsilon^0 =
  \sigma_\epsilon\tau_\epsilon\frac{\dot{r}}{f_\epsilon},\quad
  m_\epsilon^1 = \sigma_\epsilon\sqrt{f_\epsilon + \dot{r}^2},
\label{m}
\ee
where $\sigma_\epsilon := \mbox{sign}(m_\epsilon^\mu\partial_\mu r)$.

There is an important relation between $\sigma$ and $\tau$ that we are
going to derive (we omit all epsilons in the indices). The Killing
vector $\xi^\mu$ is tangential to $r =$ const curves; hence, $\xi^\mu$
is orthogonal to the vectors $\rho^\mu$ defined by $\rho^\mu :=
g^{\mu\nu}\partial_\nu r$. Thus, there are positive numbers $\xi$ and
$\rho$, almost everywhere along $\Sigma$, such that
\[
  \left( \frac{\xi^\mu}{\xi}, \frac{\rho^\mu}{\rho} \right) \quad
\mbox{or}\quad
  \left( \frac{\rho^\mu}{\rho}, \frac{\xi^\mu}{\xi} \right) 
\] 
(the timelike vector first) is an orthonormal dyad. Similarly, 
$(T^\mu, m^\mu)$ is an orthonormal dyad at the same point so that
\beann
  T^\mu & = & a \frac{\xi^\mu}{\xi} + b \frac{\rho^\mu}{\rho}, \\
  \kappa m^\mu & = & b \frac{\xi^\mu}{\xi} + a \frac{\rho^\mu}{\rho},
\eeann
where $a$ and $b$ are some reals and $\kappa = \pm 1$; if $\kappa = +1
(-1)$, the two dyads have the same (opposite) orientation. Then,
\[
  - g_{\mu\nu}T^\mu\xi^\nu = - a(\mbox{sign}\xi^\mu)\xi,
\]
\[
  \sigma = \mbox{sign}(g_{\mu\nu}m^\mu\rho^\nu) = \mbox{sign}(\kappa a
  (\mbox{sign}\rho^\mu)), 
\]
where $\mbox{sign}\,v^\mu = +1\ (-1)$, if $v^\mu$ is spacelike
(timelike). Thus,
\[
  \sigma =
  \kappa(\mbox{sign}\rho^\mu)(\mbox{sign}\xi^\mu)(-\tau) =
  \kappa \tau, 
\]
because $(\mbox{sign}\rho^\mu)(\mbox{sign}\xi^\mu) = -1$. $\kappa$ can easily
be determined. We distinguish two cases: $T^\mu$ is future or past oriented,
and introduce the corresponding sign $\eta = \pm 1$. Consider first the future
oriented case. In $Q_I$, $\xi^\mu$ and $T^\mu$ are both timelike and future
oriented, so $a > 0$. $m^\mu$ and $\rho^\mu$ are both spacelike, $\rho^\mu$ is
oriented to the right and $\epsilon m^\mu$ to the left; hence,
$\kappa_\epsilon = -\epsilon$. In $Q_{II}$, $\rho^\mu$ and $T^\mu$ are both
timelike and both future oriented, so $b > 0$. $m^\mu$ and $\xi^\mu$ are
spacelike, $\xi^\mu$ oriented right and $\epsilon m^\mu$ left; again,
$\kappa_\epsilon = -\epsilon$. A similar analysis for the remaining quadrants
confirms the relation $\sigma_\epsilon = -\epsilon\tau_\epsilon$ for future
oriented shells.  Suppose now that $T^\mu$ is past oriented. Then $\tau$ just
changes sign, but everything else (in particular, the $\sigma$'s) remains as
before; hence, $\sigma_\epsilon = \epsilon\tau_\epsilon$ for past oriented
shells. Thus, the results can be summarized by the equation 
\be
\sigma_\epsilon = -\eta\epsilon\tau_\epsilon.
\label{sigma-e}
\ee

Finally, we calculate the acceleration $a_\epsilon$ of the shell in
the spacetime $M_\epsilon$ directed out of $M_\epsilon$. It is defined
by
\[
  a_\epsilon = m_{\epsilon\,\rho}\nabla^\epsilon_sT^\rho_\epsilon,
\]
where
\[
  \nabla_s^\epsilon T^\rho_\epsilon := \dot{T}^\rho_\epsilon +
  \Gamma^\rho_{\epsilon\,\mu\nu}T^\mu_\epsilon T^\nu_\epsilon,
\]
and $\Gamma^\rho_{\epsilon\,\mu\nu}$ are the Christoffel symbols of
the Kruskal metric $g_{\epsilon\,\mu\nu}$. Substituting for
$\dot{t}_\epsilon$ from Eq.\ (\ref{e}) and using Eq.\ (\ref{m}), we
obtain after some rearrangements
\be
  a_\epsilon =
  \sigma_\epsilon\frac{(\sqrt{f_\epsilon + 
  \dot{r}^2})^{\mbox{\textbf .}}}{\dot{r}}.
\label{a}
\ee

These relations will be important for the exact counting of possible
solutions of Israel's equation, which determines the motion of the shell.

\subsection{The matter of the shell}
\label{sec:matter}
We assume the shell to be made of ideal fluid. A spherically
symmetric shell matter is ``isotropic'', and so it is always ``ideal
fluid'', though the fluid could have some internal degrees of freedom
(cf.\ \cite{B-K}), which we are suppresing. Thus, we set
\[ 
  T^{kl} = (\rho + p) T^k T^l + p\gamma^{kl},
\]
where $\gamma_{kl} := g^\epsilon_{\mu\nu}X_{\epsilon\,k}^\mu
X_{\epsilon\,l}^\nu$ is the induced metric on $\Sigma$, 
\[
  X_{\epsilon\,k}^\mu := \frac{\partial x_\epsilon^\mu}{\partial y^k},
\]
$x_\epsilon^\mu(y^0,y^1,y^2)$ are the embedding functions of $\Sigma$
in ${\mathcal M}_{1,2}$. If we choose functions (\ref{Sigma}) as
embedding functions we obtain
\[
  \gamma_{ss} = -1,\quad \gamma_{\vartheta\vartheta} =
  r^2(s),\quad \gamma_{\varphi\varphi} = r^2(s)\sin^2\vartheta;
\]
$\rho$ is the surface mass density and $p$ is the surface pressure
(negative of surface tension).  We assume that the state of the shell
matter depends only on its total surface area ${\mathcal A}:= 4\pi
r^2$ and that the kind of the material is defined by the barotropic
equation of state $p = p(\rho)$. We will also assume that the
stress-energy in the shell is conserved (which follows, if there is
vacuum outside the shell, from Israel's equation---see later).  Thus,
\[
  T^{kl}_{\ \ |l} = 0,
\]
where the vertical bar denotes the covariant derivative associated
with the metric $\gamma_{kl}$. If one substitutes for the
stress-energy tensor, one obtains the energy equation
\[
  (\rho T^k)_{|k} + pT^k_{\ |k} = 0.
\]
This is equivalent to
\be
  {\mathcal A}\frac{d\rho}{d{\mathcal A}} + \rho + p(\rho) = 0.
\label{energy}
\ee
The solution to this equation will be of the form $\rho({\mathcal A})$ and
it will depend on one constant. We define our system to be just {\em one} particular solution to
Eq.\ (\ref{energy}). We assume that the solution defines a
positive density $\rho$ for all $r \in (0,\infty)$. Thus, there is a
well-behaved unique rest mass density at any radius. In this way, the
system is as elementary as possible: it has no internal degree of
freedom and its total energy depends only on its radius and velocity.
Finally, we define the so-called {\em mass function} $M(r)$ by
\[
  M(r) := {\mathcal A}(r)\rho({\mathcal A}(r));
\]
this definition is justified by the fact that $M(r)$ will appear in
many equations at a prominent place (in some cases, it represents the
proper mass of the shell).

As an example, consider the equation of state 
\[
  p = k\rho,
\]
where $k \in (-1,+1)$. The correponding solution to Eq. (\ref{energy})
reads
\[
  \rho = m_1{\mathcal A}^{-k-1},
\]
where $m_1$ is a constant. If $k > 0$ (positive pressure), then $M(0)
= \infty$ and $M(\infty) = 0$; if $k < 0$ (positive surface tension),
then $M(0) = 0$ and $M(\infty) = \infty$; for $k = 0$ (dust), $M =$
const. 

We summarize: the non-zero components of the stress-energy tensor have
the form
\be
  T^{ss} = \rho,\quad T^{\vartheta\vartheta} =
  \frac{p(\rho)}{r^2},\quad T^{\varphi\varphi} =
  \frac{p(\rho)}{r^2\sin^2\vartheta}.
\label{T-end}
\ee

\subsection{Israel's equation}
\label{sec:israel}
The dynamics of a (massive) shell is governed by equations found by
Dautcourt \cite{dautcourt} and brought to a nice geometrical form by
Israel \cite{israel}. Israel's equation can be written in the form
\be
  Q_+^{kl} + Q_-^{kl} = 8\pi\mbox{G}T^{kl},
\label{israel}
\ee
where $Q_\epsilon^{kl} := L_\epsilon\gamma^{kl} - L_\epsilon^{kl}$ and
$L_{\epsilon\,kl} := - m_{\epsilon\,\mu;\nu}X_{\epsilon\,k}^\mu
X_{\epsilon\,l}^\nu$ is the second fundamental form of $\Sigma$ in
${\mathcal M}_{1-}$ or ${\mathcal M}_{2+}$.

The way (\ref{israel}) of writing the Israel equation makes the
independence of the equation on the orientation of the surface
$\Sigma$ manifest. It is also invariant with respect to any coordinate
change of $y$'s, in particular with respect to the time orientation
change $y^0 \mapsto -y^0$. If we use the embedding formulas (\ref{Sigma}),
we obtain 
\beann
  L^\epsilon_{ss} & = &  a_\epsilon \\
  L^\epsilon_{\vartheta\vartheta} & = & -r\sigma_\epsilon\sqrt{f_\epsilon +
  \dot{r}^2}, \\
  L^\epsilon_{\varphi\varphi} & = & -r\sigma_\epsilon\sqrt{f_\epsilon +
  \dot{r}^2}\sin^2\vartheta
\eeann
where $a_\epsilon$ is given by Eq.\ (\ref{a}). Then the trace reads
\[
  L_\epsilon = - a_\epsilon
  - 2 \sigma_\epsilon \frac{\sqrt{f_\epsilon + \dot{r}^2}}{r},
\]
and
\beann
  Q^{ss}_\epsilon & = & 2 \sigma_\epsilon \frac{\sqrt{f_\epsilon +
  \dot{r}^2}}{r}, \\
  Q^{\vartheta\vartheta}_\epsilon & = & - \frac{1}{r^2}\left(
  a_\epsilon +
  \sigma_\epsilon \frac{\sqrt{f_\epsilon + \dot{r}^2}}{r} \right), \\
  Q^{\varphi\varphi}_\epsilon & = & - \frac{1}{r^2\sin^2\vartheta}\left(
  a_\epsilon + 
  \sigma_\epsilon \frac{\sqrt{f_\epsilon + \dot{r}^2}}{r} \right).
\eeann

If we substitute these formulas into the L.H.S. and Eqs.\
(\ref{T-end}) into the R.H.S. of Eq.\ (\ref{israel}), we obtain 
Israel's equation for our model:
\bea
  \sigma_+\sqrt{f_+ + \dot{r}^2} + \sigma_-\sqrt{f_- +
  \dot{r}^2} & = & 4\pi\mbox{G}r\rho,
\label{israel1} \\
  a_+ + a_-
  + \sigma_+\frac{\sqrt{f_+ + \dot{r}^2}}{r} +
  \sigma_-\frac{\sqrt{f_- + \dot{r}^2}}{r} & = & - 8\pi\mbox{G}p.
\label{israel2}
\eea 
If we substitute for $a_\epsilon$ from Eq.\ (\ref{a}) into
(\ref{israel2}), then the resulting equation together with the time
derivative of Eq.\ (\ref{israel1}) implies the energy equation
(\ref{energy}), or the time derivative of Eq.\ (\ref{israel1})
together with Eq.\ (\ref{energy}) implies the resulting equation.

We can summarize: the full dynamics of the shell is contained in the
following equations: 
\bea
  f_\epsilon\dot{t}_\epsilon = -\eta\epsilon\sigma_\epsilon
  \sqrt{f_\epsilon + \dot{r}^2},
\label{israel-t} \\
  \sigma_+\sqrt{f_+ + \dot{r}^2} + \sigma_-\sqrt{f_- +
  \dot{r}^2} & = & \frac{\mbox{G}M(r)}{r},
\label{israel-sqrt}
\eea
where the meaning of the paremeters $\eta$, $\epsilon$,
$\sigma_\epsilon$ (all equal to $\pm 1$) was explained above,
$f_\epsilon = 1 - 2\mbox{G}E_\epsilon/r$, and $M(r)$ is the mass
function defined in Sec.\ \ref{sec:matter}.

\subsection{The radial equation}
\label{sec:radial}
Double squaring Eq.\ (\ref{israel-sqrt}), we obtain the radial
equation
\be
  \dot{r}^2 + V(r) = 0,
\label{radial}
\ee
where
\be
  V(r) = -\frac{\mbox{G}^2}{4}\frac{M^2(r)}{r^2} - \frac{\mbox{G}(E_+
  + E_-)}{r} - \frac{(E_+ - E_-)^2}{M^2(r)} + 1. 
\label{potent}
\ee
 From the point of view of the existence and properties of solutions, Eq.\
(\ref{radial}) is easier to study than Eq.\ (\ref{israel-sqrt}).
Let us do this study first and postpone the question of how the
solutions of these two equations are exactly related.

Eq.\ (\ref{radial}) is invariant with respect to the transformations
\[
  s \leftrightarrow -s,\quad \mbox{or}\quad s \mapsto s +
  s_0, 
\]
and
\[
  E_+ \leftrightarrow E_-,\quad\mbox{or}\quad M \leftrightarrow -M.
\]

Suppose that the pair $(E_-,E_+)$ is given. Eq.\ (\ref{radial})
together with the initial condition $r(0) = r_0$ has either {\em no}
solution, if $V(r_0) > 0$, or {\em two} solutions, $r(s)$ and $r(-
s)$, if $V(r_0) < 0$. Thus, for each maximal interval $(r_1,r_2)$ in
which $V(r) < 0$, we obtain just one or just two maximal solutions,
depending on what happens at the points $r_1$ and $r_2$, where $V(r_1)
= V(r_2) =0$. If $V'(r_2) \neq 0$, then we can extend the solution
$r(s)$ by $r(-s)$ at this point; similarly, if $r_1 \neq 0$ and
$V'(r_1) \neq 0$. In each of these cases, there is only one solution
in the interval. If none of these cases takes place, then we have two
maximal solutions, $r(s)$ and $r(-s)$, in the interval; they can be
chosen in some standart way for each interval.  Any other solution
within the interval can be obtained by the shift $s \mapsto s + s_0$
so that it satisfies the initial condition $r(s_0) = r_0$ for any $r_0
\in (r_1,r_2)$. The solutions that reach infinity can only exist if
$(E_1,E_2)$ is such that $V(\infty) \leq 0$.

Let us consider an example: the dust matter, which has $p = 0$ and thus
$M =$ const. For the potential (\ref{potent}) we obtain
\[
  V(r) = -\frac{\mbox{G}^2M^2}{4}\,\left(\frac{1}{r}\right)^2 -
  \mbox{G}(E_+ + E_-)\left(\frac{1}{r}\right) - \frac{(E_+ -
  E_-)^2}{M^2} + 1.  
\]
Then
\[
  \frac{dV}{dr} = \frac{\mbox{G}^2M^2}{2}\,\left(\frac{1}{r}\right)^3 +
  \mbox{G}(E_+ + E_-)\left(\frac{1}{r}\right)^2,
\]
so that 
\[
  \frac{dV}{dr} > 0
\]
in the whole range $r \in (0,\infty)$. Further,
\[
  V(0) = -\infty,\quad V(\infty) = 1 - \frac{(E_+ - E_-)^2}{M^2}.
\]
For given $E_+$ and $E_-$, there is a unique $r_m$ such that $V(r_m)
= 0$ (allowing for the value $r_m = \infty$), and the range of
possible initial values $r_0$ for which a solution exists is then
$[0,r_m]$. Hence, we always have a unique solution satisfying $r(0) =
0$, if $r_m < \infty$. There are two solutions, $r(s)$ and
$r(-s)$, satisfying $r(0) = 0$ if $r_m = \infty$. 

We shall have hyperbolic solutions if $V(\infty) < 0$, a parabolic
solution if $V(\infty) =0$, and elliptic solutions if $V(\infty) > 0$.
For the elliptic (bound) solutions, $r_m < \infty$. The unique
non-negative solution of the equation $V(r) = 0$ is
\[
  \frac{\mbox{G}}{r_m} = \frac{2}{M^2}\left(\sqrt{4E_+E_- + M^2} - E_+
  - E_-\right). 
\]
The parabolic solutions satisfy $E_+ - E_- = \pm M$. If the maximal
radius reaches a horizon, $r_m = 2\mbox{G}E_\epsilon$, the condition
\[
E_{-\epsilon} - E_\epsilon = - \frac{M^2}{4E_\epsilon}
\]
must be satisfied.

We can easily understand the spectrum of $E_+$ for a fixed $E_-$. 
Three cases can be recognized according to the value of $E_-$:

\subsubsection*{Case $E_- < M/2$}
If $E_+ \in (E_- + M, \infty)$, we have the hyperbolic solutions, if
$E_+ = E_- + M$ parabolic, and for $E_+ \in (0,E_- + M)$ elliptic. At
$E_+ = (E_- + \sqrt{E_-^2 + M^2})/2$, $r_m = 2\mbox{G}E_+$; if $E_+$
decreases beyond this value, then $r_m$ also decreases but it does not
reach $2\mbox{G}E_-$; the minimal value of $r_m$ is
$(\mbox{G}M^2)/(2(M-E_-))$. 

\subsubsection*{Case $E_- \in [M/2,M)$}
This is the same as the previous case but if $E_+$ decreases after
$r_m$ reaches the value $2\mbox{G}E_+$, $r_m$ can also reach  the value
$2\mbox{G}E_-$---at $E_+ = E_- - M^2/(4E_-)$; then, as $E_+$ decreases
further, $r_m$ increases again reaching the value
$(\mbox{G}M^2)/(2(M-E_-)) > 2\mbox{G}E_-$ at $E_+ = 0$.

\subsubsection*{Case $E_- \in [M,\infty)$}
The hyperbolic solutions fill two intervals: for $E_+ \in (E_- + M,
\infty)$ they reach the right infinity, and for $E_+ \in (0,E_- - M)$
they reach the left one. We have two parabolic solutions,
$E_+ = E_- \pm M$, and in the interval $E_+ \in (E_- - M, E_- + M)$
the solutions are elliptic.

\subsection{Relation between solutions of the radial
equation and the shell spacetime} 
\label{sec:relat}
The radial equation (\ref{radial}) is a consequence of Israel's
equation (\ref{israel-sqrt}). Hence, every solution of Eq.\
(\ref{israel-sqrt}) must solve Eq.\ (\ref{radial}). However, as 
(\ref{radial}) is obtained by squaring (\ref{israel-sqrt}) twice, some
solutions to (\ref{radial}) need not solve (\ref{israel-sqrt}): one
could loose the information about the signs $\sigma_\epsilon$. 

For the understanding of this problem, the following observation is
crucial. If one calculates the expression $f_\epsilon + \dot{r}^2$
from the radial equation (\ref{radial}), one obtains:
\be
  f_\epsilon + \dot{r}^2 = J_\epsilon^2,
\label{f+rdot}
\ee
\be
  J_\epsilon := \frac{\mbox{G}M(r)}{2r} -
  \epsilon\,\frac{E_+ - E_-}{M(r)}.
\label{J}
\ee
Thus, the expression on the left hand side of Israel's equation,
\be
  \sigma_+\sqrt{f_+ + \dot{r}^2} + \sigma_-\sqrt{f_- + \dot{r}^2}
\label{sigmas}
\ee
with some signs $\sigma_\pm$ equals to
\[
  \sigma_+ |J_+| + \sigma_- |J_-|.
\]
It follows immediately that the only choice of the signs that
satisfies Israel's equation (\ref{israel-sqrt}) is
\be
  \sigma_\epsilon = \mbox{sign}\,J_\epsilon.
\label{unique-s}
\ee
To obtain a valid shell spacetime, we have to satisfy also Eqs.\ (\ref{e})
and (\ref{sigma-e}); this leads to
\be
  f_\epsilon(r(s))\dot{t}_\epsilon = - \eta\epsilon\,J_\epsilon.
\label{time-eq}
\ee
Thus, quite surprisingly, each solution of the radial
equation (\ref{radial}) generates a solution of Eqs.\ (\ref{israel-t})
and (\ref{israel-sqrt}) which describe the full dynamics of the shell. 

Let us construct a shell spacetime with a
future oriented shell motion from a solution of the radial equation:
\ben
\item Choose an oriented pair $(E_1,E_2)$ of positive numbers; this
defines two Kruskal spacetimes, and we will construct a shell
spacetime by having ${\mathcal M}_1$ with $E_1 = E_-$ to the left and
${\mathcal M}_2$ with $E_2 = E_+$ to the right.
\item Choose a solution $r(s)$ to the radial equation.
Then the expressions $\sigma_\epsilon$ given by Eq.\ (\ref{unique-s}) 
are well-defined for each $s$ and $\epsilon$. The function
$r(s)$ solves then Eq.\ (\ref{israel-sqrt}) with the signs
$\sigma_\epsilon$. 
\item Eq.\ (\ref{time-eq}) determines the functions
$t_\epsilon(s)$ up
to an additive constant, $t_\epsilon(s) \mapsto t_\epsilon(s) +
c_\epsilon$. 
\item The pair of functions $r(s), t_\epsilon(s)$ given on some
common interval of $s$ determines two different curves in each
Kruskal spacetime ${\mathcal M}_1$ and ${\mathcal M}_2$. They are related by
the inversion map I (cf.\ Sec.\ \ref{sec:kruskal}), so
only one of them is future oriented. Thus, in each of ${\mathcal M}_1$ and
${\mathcal M}_2$, there is a unique surface $\Sigma_1$ and $\Sigma_2$,
respectively. We cut the spacetimes along it and solder the
parts together. The geometry of the resulting spacetime does not
depend on $c_\epsilon$: the constants can be made zero by
continuous isometries in each half.
\een

The construction also revealed some ambiguities. If we did not fix the
orientation of the shell movement and the order of the pair
$(E_-,E_+)$, we would have obtained four non isometrical spacetimes
in general. This corresponds to the number of elements in the discrete
subgroup of the Kruskal group $G_K$. Reversing the order of
$(E_-,E_+)$ corresponds to the map C (cf.\ Sec.\ \ref{sec:kruskal})
performed in each spacetime $({\mathcal M}_1,E_1)$ and $({\mathcal
  M}_2,E_2)$---the dependence of $r$ on $s$ along any 
curve cannot be changed by such a map. Similarly, if we allowed also
for past oriented shells, then the two functions $t_\epsilon(s)$ and
$r(s)$ would determine two curves in each spacetime that are related
by the map I; by inverting the parameter $s \mapsto -s$, we obtain a
future-oriented shell motion. Let us denote the parameter inversion by
T$_s$. It is clear from this discussion that a description of the
dynamics that is based solely on the radial trajectory $r(s)$ will not
distinguish physically different states. More concretely: the radial
equation (\ref{radial}) is quadratic in the total energy $E_+$; if
considered as an equation for $E_+$, it has two roots in
general. Hence, $E_+$ is not a well-defined function on the space of
allowed pairs $(r,\dot{r})$.

In fact, there are even more states to be distinguished: sometimes, two
spacetimes that are otherwise completely isometric to each other can
describe different physical states. The crucial point is whether the
isometry moves also the observers or not. Let us assume that the
observers in our model are assembled at the right spacelike infinity.
If we move the shell by the continuous subgroup of $G_K$, $t_+
\mapsto t_+ + c_+$, {\em without} moving the observers, then the physical
state will change: the observer will be able to detect $c_+$ as a
change of the shell arrival time at a fixed radius, for example. Thus,
one and the same radial trajectory $r(s)$ represents $4\times\infty$
different physical states (or $2\times\infty$, if one does not like
the past-oriented shells). Observe that the time shift $t_- \mapsto
t_- + c_-$ does not do anything to change the physical state with
respect to the right family of observers. Thus, we have to couple the
internal and the external times, if we wish to describe the shell
dynamics by means of variables that refer to the left (internal)
spacetime.

Another important point is that the validity of our equations is not
limited to spacetimes with just one shell; as far as there are no
crossings, there can be other shells and/or sources.  Interesting
examples of two-shell spacetimes can be easily constructed using the
symmetries. Let us take three Kruskal spacetimes, $({\mathcal M}_1,E_1)$,
$({\mathcal M}_2,E_2)$, and $({\mathcal M}_3,E_3)$ such that $E_1 = E_3$. The
shell $\Sigma_{12}$ between ${\mathcal M}_1$ and ${\mathcal M}_2$ can be taken
as the C- or I$\circ \mbox{T}_s$-image of the shell $\Sigma_{23}$ between
${\mathcal M}_2$ and ${\mathcal M}_3$. Consider a source lying far left in
$({\mathcal M}_1,E_1)$ so that both shells are to the right from it. Such a
source can be constructed explicitly by taking for example a suitable
Oppenheimer-Snyder star. The motion of both shells is not changed by this
source. We can interpret this as follows. The difference $e_{12} := E_2 - E_1$
is the contribution of the energy of the shell $\Sigma_{12}$ to the total
energy of the system, and similarly $e_{23} := E_3 - E_2$ for
$\Sigma_{23}$; these contributions exactly cancel each other:
\[
  e_{12} = - e_{23}.
\]
One can consider $e$ as the {\em gravitational charge} of the shell.
Then $\Sigma_{12}$ and $\Sigma_{23}$ have opposite gravitational
charges. It is interesting to observe that the opposite charges give
rise to opposite accellerations.  Let us calculate the acceleration
$a_{1-}$ of $\Sigma_{12}$ in ${\mathcal M}_1$ directed out of ${\mathcal M}_1$
and the acceleration $-a_{3+}$ of $\Sigma_{23}$ in ${\mathcal M}_3$ directed
into ${\mathcal M}_3$ 
(both pointing right). Using the formulas (\ref{a}), (\ref{f+rdot}),
(\ref{J}) and (\ref{unique-s}), we obtain an important formula for
$a_\epsilon$:
\[
  a_\epsilon = \frac{d}{dr}J_\epsilon.
\]
Substituting $e_{12}$  or $e_{23}$, respectively, for $E_+ - E_-$ in
Eq.\ (\ref{J}) we
obtain:
\beann
 a_{1-} & = & \frac{d}{dr}\left(\frac{\mbox{G}M(r)}{2r} +
 \frac{e_{12}}{M(r)}\right) \\
 - a_{3+} & = &  - \frac{d}{dr}\left(\frac{\mbox{G}M(r)}{2r} -
 \frac{e_{23}}{M(r)}\right).
\eeann
However, $e_{12} = - e_{23}$, so $a_{1-} = -(-a_{3+})$ as claimed.

An interesting example is the following. Let $\Sigma_{23}$ be a
hyperbolic shell outgoing to the right of the event horizons of both
spacetimes $({\mathcal M}_2,E_2)$ and $({\mathcal M}_3,E_3)$; it will reach
the right assymptotic region sometimes in future. The other shell,
$\Sigma_{12}$, can be constructed by the map I$\circ\mbox{T}_s$ of the
first one in $({\mathcal M}_2,E_2)$ and $({\mathcal M}_3,E_3)$, where the map
I is interpreted as a map I : ${\mathcal M}_3 \mapsto {\mathcal M}_1$. Thus,
$\Sigma_{12}$ is a 
hyperbolic outgoing (with respect to the right infinity) state lying
left from the event horizons of both $({\mathcal M}_1,E_1)$ and $({\mathcal
  M}_2,E_2)$. If 
we could construct a quantum theory in such a way that these will be a
pair of shells whose classical interpretation will coincide with our
construction, and if we can get creation of such pairs, spontaneous or
induced by some agent, then this may give a model of Hawking effect
with some sort of back reaction (cf.\ \cite{K-W}). There seem to be
other possibilities, too: $\Sigma_{12}$ can be constructed from
$\Sigma_{23}$ by C. One should try to implement these speculations by
a suitable quantization of the system.

\section{Null shells}
\label{sec:null}
The spherically symmetric shell from null matter moving in vacuum has a very
simple theory \cite{D-tH}, \cite{B-I}. Its motion with respect to each
of the embedding spacetimes ${\mathcal M}_1$ and ${\mathcal M}_2$ is that of
radial null geodesics. Thus, if $x^a_\epsilon$, $a=0,1$, are two coordinates in
the Kruskal $uv$-plane, and $g_{\epsilon\,ab}(x)$ the corresponding
metric, the dynamical equation of the shell takes the form
\[
g^\epsilon_{ab}\dot{x}^a_\epsilon\dot{x}^b_\epsilon = 0,
\]
where the dot is the derivative with respect to an arbitrary parameter
(in a two-dimensional spacetime, every smooth null curve is a null
geodesic). In local Schwarz\-schild coordinates (\ref{Smetric}) we obtain
\be
  -f_\epsilon\dot{t}^2_\epsilon + f^{-1}_\epsilon\dot{r}^2 = 0.
\label{null-dyn}
\ee 
There are only six different types of null geodesics in any
two-dimensional Kruskal spacetime (we consider only the future
oriented shells): 
\ben
\item generic geodesics, that reach the right (left) asymptotic region;
we distinguish them by the sign parameter $\zeta = +1$ ($\zeta = -1$).
\item outgoing ($\zeta\dot{r} > 0$) and ingoing ($\zeta\dot{r} < 0$)
for each $\zeta$-type (out- and ingoing is defined with respect to
the right infinity).
\item outgoing and ingoing horizons.
\een
Along any generic null geodesics, the coordinate $r$ is an affine parameter
and it behaves  strictly monotonically, acquiring all values from the
interval $(0,\infty)$. There are only two exceptions: the horizons.

Given two generic null geodesics of the same type, one in ${\mathcal M}_1$ and
the other in ${\mathcal M}_2$, they define the surfaces $\Sigma_1$ and
$\Sigma_2$, and there is only one soldering of ${\mathcal M}_{1-}$ with
${\mathcal M}_{2+}$
along $\Sigma_1$ and $\Sigma_2$: the function $r$ must be continuous.
This is possible for any values of $E_-$ and $E_+$. In particular,
the outgoing (ingoing) horizon in ${\mathcal M}_1$ can be soldered only to the
outgoing (ingoing) horizon in ${\mathcal M}_2$, and this only if $E_1 = E_2$.
However, for a null dust shell (which is pressureless), the soldering
must be affine (the shell is ``affinely conciliable''---cf.\
\cite{B-I}), and this 
is not unique. For the outgoing horizon, using the double null
Kruskal coordinates $u$ and $v$, a general soldering is given by
$u_- = 0$, $u_+ = 0$, $v_+ = v_- + \alpha$, where $\alpha$ is an
arbitrary constant, and similarly for the ingoing case: $v_- = v_+ =
0$, $u_+ = u_- + \beta$, $\beta$ arbitrary.

It follows that we can define the quantity $e_\lambda$ for all affine
parameters $\lambda$ along any null shell. We define first
$e_{\lambda\epsilon}$ by
\[
  e_{\lambda\epsilon} := \frac{dx^a_\epsilon}{d\lambda}\xi_{\epsilon\,a},
\]
where $\xi^a_\epsilon$ is the Killing vector (\ref{killing}) of the
respective Kruskal spacetime. As $x^a_\epsilon(\lambda)$ is a null
geodesic in each of the Kruskal spacetimes, $e_{\lambda\epsilon}$ is
conserved along the curve on both sides. If the shell reaches an
asymptotic region, both geodesics reach the same (right or left)
asymptotic region in the respective spacetime, and so both
$e_{\lambda+}$ and $e_{\lambda-}$ have the same sign. From Eq.\
(\ref{null-dyn}), we obtain
\be
  \left(f_\epsilon\frac{dt_\epsilon}{d\lambda}\right)^2 =
  \left(\frac{dr}{d\lambda}\right)^2;
\label{dr/dl}
\ee
but, in the Schwarzschild coordinates,
\be
  e_{\lambda\epsilon} = f_\epsilon\frac{dt_\epsilon}{d\lambda}.
\label{e-lambda}
\ee
Hence, $e_{\lambda+} = e_{\lambda-}$. Moreover, if $\lambda = r$, then
$e_{\lambda\epsilon} = \zeta$. For null shells, we thus have 
$e_{\lambda-} = e_{\lambda+}$, and we can define
\[
  e_\lambda := e_{\lambda-} = e_{\lambda+}.
\]

The structure of matter of the given null shell constructed
according to the recipe above is given by the  Barrab\`{e}s-Israel
equations. In particular, for a generic shell, Eqs.\ (52) and (53) of
\cite{B-I} imply

1) the pressure $p$ inside the shell vanishes (we have null dust if
the shell is surrounded by vacuum),

2) the shell stress-energy tensor $T^{ab}$ is given by
\be
  T^{ab} = \zeta\frac{E_+ - E_-}{4\pi r^2}\,l^al^b,
\label{null-T}
\ee
where $l^a$ is the tangent vector to the radial null geodesics
affinely parametrized by $r$; in the shell coordinates $y^k$,
\[
  l^k = \frac{\partial y^k}{\partial r}.
\]

For a horizon shell (see \cite{D-tH}), the only non-vanishing
components of the stress-energy tensor with respect to the coordinates
$(v,\vartheta,\varphi)$ (or $(u,\vartheta,\varphi)$) are
\be
  T^{vv} = \frac{\mbox{e}}{(32\mbox{G}E)^2\pi}\alpha,\quad
  (T^{uu} = \frac{\mbox{e}}{(32\mbox{G}E)^2\pi}\beta),
\label{T-horiz}
\ee
where e is the basis of natural logarithms and $E = E_- = E_+$.

The energy density of the shell is a measurable quantity (see
\cite{B-I}). In order that this be positive, the condition
\be
  E_+ - E_- = \zeta|E_+ - E_-|
\label{sign-E}
\ee
must hold for generic shells, and some conditions on the sign of
$\alpha$ and $\beta$ for the horizon shells.

 From the formulas (\ref{null-dyn}), (\ref{null-T}) and
(\ref{T-horiz}) it follows that the matter content of the null shell
is independent of their trajectories; they seem to have an internal
degree of freedom. However, it is possible, at least formally, to
create a relation between $\dot{r}$ and $E_+ - E_-$ by hand, using the
arbitrariness of an affine parameter $\lambda$. This will turn out to
be necessary in the Hamiltonian formalism where the final
justification for this step lies.

Let us define: the affine parameter $\lambda$ along shells is a {\em
physical parameter}, if
\be
  f_\epsilon\frac{dt_\epsilon}{d\lambda} = \eta(E_+ -E_-)
\label{phys-p}
\ee
where, as in the case of massive shells, $\eta = +1 (-1)$ if $l^a$ is
future (past) oriented.
(This also keeps the mass density positive for generic shells.) Then,
Eqs.\ (\ref{dr/dl}) and (\ref{e-lambda}) imply
\be
  \left(\frac{dr}{d\lambda}\right)^2 = (E_+ - E_-)^2.
\label{rad-null}
\ee
This is the ``radial equation for the null shells'' corresponding to
the Eqs.\ (\ref{radial}) and (\ref{potent}) for massive
shells. Together with the time equation 
\be
  f_\epsilon\frac{dt_\epsilon}{d\lambda} = \eta\zeta
  \left|\frac{dr}{d\lambda}\right|,
\label{time-null}
\ee
it determines the trajectory (the left hand side does not depend on
$\epsilon$). 

In this form, the null-shell dynamics is just a limit case of the
massive-shell dynamics (with the exception of the horizon shells). Indeed,
let us define the physical parameter $\lambda$ for massive shells by
\[
  \frac{ds}{d\lambda} = M(r(s)).
\]
Then, the radial equation (\ref{radial}) and (\ref{potent}) becomes
\be
  \left(\frac{dr}{d\lambda}\right)^2 + \tilde{V}(r) = 0,
\label{rad-lambda}
\ee
where
\be
  \tilde{V}(r) = -\frac{\mbox{G}^2M^4(r)}{4r^2} -
  \frac{\mbox{G}M^2(r)(E_+ + E_-)}{r} - (E_+ - E_-)^2 + M^2(r).
\label{potennt-lambda}
\ee
Clearly, $\lim_{M(r)=0}\tilde{V}(r) = - (E_+ - E_-)^2$, and Eq.\
(\ref{rad-lambda}) goes over into Eq.\ (\ref{rad-null}).

\section{Shells as particles on two-dimensional Kruskal spacetimes}
\label{sec:particle}
In this section, we will try to rewrite the dynamics of the shells 
as a dynamics of a fictitious particle on a {\em fixed}
two-dimensional Kruskal background. 

Let us study the motion of a point particle on a two-dimensional Kruskal
spacetime with the energy $E$ that results from the action
\be
  S = \int ds(p_0\dot{x}^0 + p_1\dot{x}^1 - {\mathcal{NH}}),
\label{action-gen}
\ee
where $x^a$ are some coordinates on the spacetime, which serve
as canonical coordinates of the particle, $p_a$ are the conjugate
momenta, $\mathcal{N}$ is a Lagrange multiplier,
\be
  {\mathcal{H}} = \frac{1}{2}\left[g^{ab}(p_a - U\xi_a)(p_b - U\xi_b) +
  W^2\right]
\label{hamilt}
\ee
is the super-Hamiltonian,
$g^{ab}(x)$ is the Kruskal metric in the coordinates $x^a$, $U(r(x))$
and $W(r(x))$ are some functions. Eq.\ (\ref{hamilt}) provides the
most general (up to an overall factor that is independent of the momenta)
super-Ha\-milton\-ian ${\mathcal H}$ with the properties 
\ben
\item $\mathcal H$ is a polynomial of second order in momenta,
\item the momentum dependence of $\mathcal H$ has the correct light-cone
structure,
\item $\mathcal H$ as a function of $x^\mu$ and $p_\mu$ is invariant with
respect to $G_K$.
\een

The action is clearly invariant with respect to all transformations of
coordinates $x^a$.

Within each quadrant of the Kruskal spacetime, we can transform the
canonical coordinates to Schwarz\-schild coordinates $t$ and $r$. The
resulting action is
\be
  S = \int ds (p_t\dot{t} + p_r\dot{r} - {\mathcal{NH}}),
\label{action-S}
\ee
where
\be
  {\mathcal H} = \frac{1}{2}\left[ - f^{-1}(p_t + Uf)^2 + fp_r^2 +
  W^2\right],
\label{hamilt-S}
\ee
and $f(r) := 1 - (2\mbox{G}E)/r$.

The action (\ref{action-S}) is meaningful only inside the four quadrants
but can be considered as a result of a canonical transformation from
the variables $x^a$, which can be global Kruskal coordinates $u$ and
$v$. The solutions can be matched through the horizons and
the matching is well-defined by the tranformation to the
Kruskal or Eddington-Finkelstein coordinates.

Variation with respect to $p_t$ and $p_r$ yields:
\beann
  f\dot{t} & = & - {\mathcal N}(p_t + Uf), \\
  \dot{r} & = & {\mathcal N}fp_r.
\eeann
We suppose first that $W \neq 0$ and choose the gauge ${\mathcal N} =
\eta/W$, where the sign $\eta$ corresponds, as before, to different time
orientations of the shell; then 
\bea
  p_t + Uf & = & -\eta Wf\dot{t}, 
\label{pt} \\
  p_r & = & \eta Wf^{-1}\dot{r},
\label{pr}
\eea
and the constraint (following from the variation of the action with
respect to ${\mathcal N}$)
\be
  {\mathcal H} = 0
\label{constraint}
\ee
 implies that the parameter $s$ is a
proper time. Eqs.\ (\ref{pt})--(\ref{constraint}), and the equation
that results from the variation of the action with respect to $t$,
\be
  p_t = \mbox{const},
\label{conserv}
\ee
form a complete system of equations of motion. The equation that is
obtained by varying the action with respect to the variable $r$ is a
consequence of the system (\ref{pt})--(\ref{conserv}). Inserting Eqs.\
(\ref{pr}) and (\ref{conserv}) into Eq.\ (\ref{constraint}), we obtain
\be
  f + \dot{r}^2 = \left(\frac{p_t + Uf}{W}\right)^2.
\label{rad-ham}
\ee 
It is easy to see that Eqs.\ (\ref{rad-ham}) and (\ref{pt}) are
equivalent to the radial equations (\ref{f+rdot}), (\ref{J}) and the
time equation (\ref{time-eq}), provided that the constants $E$ and
$p_t$ are chosen properly.  Let $E = E_-$. Then Eq.\ (\ref{f+rdot})
with $\epsilon = -1$ will be satisfied if 
\bea 
  p_t & = & - E_+,
\label{choice-pt} \\
  Uf_- & = & E_- - \frac{\mbox{G}M^2(r)}{2r},
\label{choice-U} \\
  W & = & M(r).
\label{choice-W}
\eea
The apparent singularity in the super-Hamiltonian ${\mathcal H}$ can be removed
if ${\mathcal H}$ is multiplied by $f(r)$ because $g^{\mu\nu}\xi_\mu\xi_\nu$
provides another factor $f(r)$ at $U^2$.

There are two important elements in the choices
(\ref{choice-pt})--(\ref{choice-W}): 
\ben
\item The choice $E = E_-$: this makes the shell to move on the left
spacetime and thus {\em decouples} the background along which the shell
moves from the gravitational field the shell produces. A great
simplification is a consequence. 
\item Eq.\ (\ref{choice-pt}): we require that the total energy of the
full system, which is $E_+$, coincides with this conserved quantity.
\een

The above considerations enable us now to formulate the following
theorem. 
\begin{thm} There is one-to-one correspondence
between the states of the motion of the shell with mass function $M(r)$
and the dynamical trajectories of a point particle with $p_t < 0$
which follow from the variational principle with the action 
\be
  S = \int ds (p_a\dot{x}^a - {\mathcal{NH}}),
\label{f-action}
\ee
where
\be
  {\mathcal H} = \frac{1}{2}\left[g_-^{ab}(p_a - U_-(r)\xi_a)(p_b -
  U_-(r)\xi_b) + M^2(r)\right],
\label{f-hamilt}
\ee
$g_-^{ab}(x)$ is the Kruskal metric with $E_-$ as energy parameter,
and 
\be
  U_-(r) = \frac{1}{f_-}\left(E_- - \frac{GM^2(r)}{2r} \right).
\label{f-U}
\ee
\end{thm}

{\bf Proof}\ \ The construction of the shell spacetime from a given
solution of the 
variational principle (\ref{f-action}) goes as follows. The
super-Hamiltonian is chosen such that the solution is identical with
the shell trajectory in the spacetime $({\mathcal M},E) = ({\mathcal
  M}_1,E_-)$. Then, setting $E_+ = - p_t$, the spacetime $({\mathcal
  M}_2,E_+)$ is well-defined and we 
have to find the shell trajectory there. The function $r(s)$ is the
same as in ${\mathcal M}_1$, but $t_+(s)$ must satisfy
\[
  f_+\dot{t}_+ = -\eta\left(\frac{\mbox{G}M(r)}{2r} - \frac{E_+ -
  E_-}{M(r)}\right). 
\]
By this the trajectory in ${\mathcal M}_2$ is fixed up to the map given by the
discrete isometry C and a map given by any element of the component of
identity of the Kruskal group $G_K$ (see Sec.\ \ref{sec:kruskal}). The
map by C does not change the resulting spacetime because always only
one and the same half ${\mathcal M}_{2+}$ of ${\mathcal M}_2$ can be soldered
to ${\mathcal M}_{1-}$. The time shift can be fixed by some coupling of the
internal time $t_-$ to the external time $t_+$. For example, we can require
that the point of the trajectory in ${\mathcal M}_1$ with $t_- = 0$ coincides
with the point of the trajectory in ${\mathcal M}_2$ with $t_+ = 0$. (Another
example is given in the next section.) We thus obtain a unique state
of the shell. On the other hand, given a shell spacetime, the
trajectory of the shell in $({\mathcal M}_1,E_-)$ is well-defined and it is a
solution of the variational principle (\ref{f-action}). There is
indeed a one-to-one correspondence, Q.E.D.

Let us show that the variational principle
(\ref{f-action}) yields also the dynamics of the null shells. If we set
$M(r) \equiv 0$ in Eqs.\ (\ref{f-hamilt}) and (\ref{f-U}), we obtain
for the super-Hamiltonian in Schwarzschild coordinates
\be
  {\mathcal H} = \frac{1}{2}[-f_-^{-1}(p_t + E_-)^2 + f_-p_r^2].
\label{null-superh}
\ee
Let us choose the gauge ${\mathcal N} = \eta$. Then variations with
respect to $p_t$ and $p_r$ yield
\bea
  f_-\dot{t}_- & = & - \eta(p_t + E_-),
\label{time-nH} \\
  \dot{r} & = & \eta f_-p_r. \nn
\eea
Inserting this into the constraint ${\mathcal H} = 0$, we obtain
\[
  -f_-\dot{t}_-^2 + f_-^{-1}\dot{r}^2 = 0.
\]
Thus, we have a null geodesic. Moreover, the comparison of Eqs.\
(\ref{time-nH}) with Eq.\ (\ref{phys-p}) shows that $s$ is the physical
parameter now. The trajectory of the shell in $({\mathcal M}_2,E_+ = -p_t)$ is
determined similarly as for the massive case; the soldering, however,
is unique only for the generic shells. For horizon shells, an affine
shift stays undetermined; and only this shift determines the spacetime and
the energy density of the shell (cf.\ Eq.\ (\ref{T-horiz})).

What is the use of the super-Hamiltonian (\ref{f-hamilt})? The
discussion of Sec.\ \ref{sec:hamilt} allows only the following
accurate statement. Any true Hamiltonian $H$ that results from
${\mathcal H}$ by some reduction procedure is a meaningful one, if its
value coincides with the total Energy of the system (that is with
$E_+$). Those reduction procedures of the constraint system
(\ref{f-action}) that result in a different value of $H$ are not
allowed.

The things have been arranged in such a way that the allowed reduction
procedures are for example those which use a stationary
foliation. That is, given any spacelike surface $\mathcal S$ of the
auxiliary Kruskal spacetime, the stationary foliation is defined as
the family $\{g(s){\mathcal S}\,|\,s \in {\mathbf R}\}$, where $g(s)$
is an element of the component of identity of the Kruskal group
$G_K$. By these foliations, however, not all allowed reduction
procedures are exhausted, as an example in the next section will show.

\section{Out-Side Story}
\label{sec:out}
In this section, we shall transform the dynamics of the shell to the
form of particle dynamics on the {\em external} spacetime
$({\mathcal M}_2,E_+)$. Such schemes are in use but they are complicated. Our
aim is to show that they can be obtained from the variational principle
(\ref{f-action}) by a reduction procedure including a choice of
gauge. We shall restrict ourselves to the null shells. As an example,
we shall rederive the Kraus-Wilczek Hamiltonian \cite{K-W} from our
super-Hamiltonian.

The physical phase space (whose points are maximal classical
solutions) of the generic null shell consists of four (topologically
separated) sectors.\footnote{The reader is referred to Ref.\
\cite{margu} and Ref.\ \cite{ADM} for some reviews of the constrained
dynamics of reparametrization invariant systems in which the concepts
we use in this section are described in detail.} The subsequent
analysis applies to the sector of those shells that reach the right
infinity and are outgoing. The other sectors can be dealt with in a
completely analogous manner.

To describe the dynamics of the outgoing shells, we choose the 
Eddington-Fin\-kel\-stein retarded coordinates $u$ and $r$. The metric
in the internal spacetime reads
\[
  ds^2 = - f_-du^2 - 2 du dr + r^2 d\Omega^2.
\]
In 
these coordinates the super-Hamiltonian (\ref{f-hamilt}) with $M(r)
\equiv 0$ is given by
\[
  {\mathcal H} = -(p_u + E_-)(p_r + f_-^{-1}E_-) + \frac{1}{2}f_-(p_r +
  f_-^{-1}E_-)^2,
\]
but we allow only the solutions with $p_r = - f_-^{-1}E_-$ as the
others, with $p_u = (f_-p_r - E_-)/2$, belong to the ingoing sectors.
Thus, we can simplify the super-Hamiltonian to $p_r + f_-^{-1}E_-$.
The equations of motion then read
\[
  \dot{u} = 0,\quad \dot{r} = {\mathcal N},\quad \dot{p}_u = 0,\quad p_r =
  - f_-^{-1}E_-. 
\]
The phase functions $u$ and $p_u$ are thus perennials (constants of
motion) spanning the physical phase space $\bar{\Gamma}$ in this
sector (cf.\ \cite{margu}); on $\bar{\Gamma}$, the perennials have the
ranges
\[
  u \in (-\infty, \infty),\quad p_u \in (-\infty,0).
\]
The constraint surface $\Gamma$ is the submanifold $p_r = -
f_-^{-1}E_-$ of the extended phase space $\tilde{\Gamma}$ with the
coordinates $r$, $p_r$, $u$ and $p_u$. Thus, we can cover the
constraint surface $\Gamma$ by
the coordinates $r$, $u$ and $p_u$ ($r \in (0,\infty)$).

The reparametrization invariant system defined above is to be reduced
so that a ``true'' Hamiltonian results. For that purpose, first a
family of transversal surfaces $\{\Gamma_t\}$---the so-called time
levels---is needed (see \cite{margu} for more detail): $\Gamma_t
\subset \Gamma$ for each $t$, and $\Gamma_t$ have exactly one
(transversal) intersection with each c-orbit (trajectory) given by
$u=$ const, $p_u =$ const, $r = \lambda$. (Here $t$ is any suitable
time parameter---not necessarily the Schwarzschild time). $\Gamma_t$
can be defined by an equation of the form
\[
  t = T(r,u,p_u),
\]
where $T(x,y,z)$ is a suitable function of the three variables. $\Gamma_t$
is a symplectic space for each $t$; as a canonical chart on
$\Gamma_t$, $(u|_{\Gamma_t}, p_u|_{\Gamma_t})$ can be chosen. Second,
for a construction of a Hamiltonian dynamics, a family of time shifts
$\theta_{tt'} : \Gamma_{t'} \mapsto \Gamma_t$ is necessary such that
each $\theta_{tt'}$ is a symplectic diffeomorphism
\cite{margu}. Clearly, in our case a trivial family of $\theta_{tt'}$
is obtained by mapping the points $(u,p_u) 
\in \Gamma_{t'}$ to $(u,p_u) \in \Gamma_{t}$; this would give a
trivial (frozen) dynamics. A more suitable possibility is to employ
the momentum $p_u$ as a generator of time shifts. Since
\[
  \{r,p_u\} = 0,\quad \{u,p_u\} = 1,
\]
and $(r,u)$ can be considered as Eddington-Finkelstein coordinates,
$p_u$ generates just the 
Schwarzschild time shift. The requirement that $\Gamma_{t+\delta}$ is
obtained from $\Gamma_t$ by $p_u\delta$ leads to the following
condition on $T$:
\[
  \{T,p_u\delta\} = \delta \quad \forall \delta \in {\mathbf R},
\]
or
\[
\frac{\partial T}{\partial u} = 1.
\]
The general solution has the form
\be
  T(r,u,p_u) = u + \tilde{T}(r,p_u).
\label{T-form}
\ee

The next requirement we impose is that the transversal surfaces can be
interpreted as time-surfaces in the external (right) spacetime. To
find such surfaces, we have to couple the external time to the
internal one in such a way that the time in $({\mathcal M}_{2+},E_+)$ can be
expressed as a function of our phase space variables. The most
straightforward and useful way is to require that the retarded time be
continuous across the shell (this can easily be achieved since it is
determined up to a constant),
\be
  u = u_- = u_+.
\label{t-coupl}
\ee
Then, using $E_+ = -p_u$, we can specify the function $T$. For
example, if we want to employ the surfaces of constant Schwarzschild
time $t_+$, then we use the formula expressing $t_+$ in terms of $u_+$
and $r$,
\be
  t_+ = u_+ + r + A\ln\left|\frac{r-A}{A}\right|,
\label{ret-t}
\ee
where $A:= -2\mbox{G}p_u$. Eq.\ (\ref{t-coupl}) then yields the
following form of $T$ (satisfying the condition (\ref{T-form})):
\[
  T(r,u,p_u) = u + r + A\ln\left|\frac{r-A}{A}\right|.
\]

Finally, instead of the perennials $u$ and $p_u$, we would like to
write our dynamics by means of the radial variable $r$ and a
corresponding conjugate momentum $p$. Let us study this transformation
for general $\tilde{T}$. Of course, $\tilde{T}$ must be such that the
function $r$ is a non-trivial coordinate along $\Gamma_t$, that is,
there must exist a solution $r = R(t-u,p_u)$ of the equation $t = u +
\tilde{T}(r,p_u)$. Thus, we have the identity
\be
  t - u = \tilde{T}(R(t-u,p_u),p_u).
\label{id-t}
\ee
To find the momentum $p(t,u,p_u)$ canonically conjugate to the
coordinate $r = R(t-u,p_u)$, we look for a canonical transformation
(for each $t$)
\[
  (u,p_u) \mapsto (r,p).
\]
The transformation will be canonical, if
\be
  \frac{\partial R}{\partial u}\frac{\partial p}{\partial p_u} -
  \frac{\partial R}{\partial p_u}\frac{\partial p}{\partial u} = 1.
\label{canon}
\ee
Let us search for $p$ in the form
\[
  p = P(R(t-u,p_u),p_u).
\]
Then Eq.\ (\ref{canon}) is equivalent to
\[
  \left(\frac{\partial P}{\partial p_u}\right)_r =
  \left(\frac{\partial R}{\partial u}\right)^{-1}_{u= t-\tilde{T}(r,p_u)},
\]
  From the identity (\ref{id-t}), we obtain the final result for the
momentum conjugate to $r$ in the form
\be
  P(r,p_u) = P_0(r) - \int dp_u
  \frac{\partial \tilde{T}(r,p_u)}{\partial r},
\label{canon-f}
\ee
where $P_0(r)$ is an arbitrary function of $r$.

{\em The Hamiltonian of Kraus and Wilczek}.
Let us use the formula (\ref{canon-f}) to rederive the
Kraus-Wilczek Hamiltonian obtained in \cite{K-W}. Kraus and Wilczek
employ the system of coordinates $\tau$ and $r$ in the external
Kruskal spacetime $({\mathcal M}_2,E_+)$ that has been defined in Sec.\
\ref{sec:hamilt} (formula (\ref{KW-t})).  The relations (\ref{ret-t})
and (\ref{KW-t}) imply that the retarded time $u_+$ can be expressed
by means of $\tau$ as follows:
\[
  u_+ = \tau - r - 2\sqrt{Ar} -
  2A\ln\left|\sqrt{\frac{r}{A}} - 1\right|. 
\]
Thus, the transversal surfaces are defined by the function $T$ of the
form:
\[
  T(r,u,p_u) = u + r + 2\sqrt{Ar} + 2A\ln\left|\sqrt{\frac{r}{A}} -
  1\right|.
\]
The function $T$ satisfies the condition (\ref{T-form}) because the
foliation of the external spacetime $({\mathcal M}_2,E_+)$ by
the Kraus-Wilczek time $\tau$ is stationary. Comparing the last
relation with Eq.\ (\ref{T-form}), we find $\tilde{T}$, from which we obtain 
\[
  \frac{\partial\tilde{T}}{\partial
  r} = \frac{1}{1 - \sqrt{\frac{A}{r}}}.
\]
Hence, the formula (\ref{canon-f}) becomes identical with Eq.\
(\ref{mom-KW}) with $\epsilon = +1$ and $-p_u = H$.

We can conclude: the non-locality of the Hamiltonian of Kraus and
Wilczek is not due to any physical property of the system, but to the
choice of gauge and to the reduction procedure.

\subsection*{Acknowledgements}
Helpful discussions with D.~Giulini, G.~Lavrelashvili, K.~V.~Kucha\v{r} and
J.~Louko are gratefully acknowledged. P.H. is thankful to National Science
Foundation grant PHY9507719, to The Tomalla Foundation, Zurich, to The
Swiss Nationalfonds and to the Max-Planck-Institut f\"{u}r
Gravitationsphysik, Potsdam for support; J.~B. is grateful to the
Institue for Theoretical Physics, University of Berne, for the kind
hospitality, and to the grants Nos. GACR-202/96/0206 and GAUK-230/1996
of the Czech Republic and the Charles University for a partial
support.


\begin{thebibliography}{99}
\bibitem{bubbles} For the treatment investigating spherical vacuum
  bubles, see, for example, Refs.\ \cite{BGG} and \cite{FGG}; K.~Lake, Phys.\
  Rev.\ 
  \textbf{D29} (1984) 1861; K.~Lake and Wevrick, Canad.\ J. Phys.\
  \textbf{64} (1986) 165; P.~Laguna-Castillo and R.~A.~Matzner, Phys.\
  Rev.\ \textbf{D34} (1986) 2913; A.~Aurilia, R.~S.~Kissack, R.~Mann and
  E.~Spallucci, Phys.\ Rev.\ \textbf{D35} (1987) 2961; V.~A.~Berezin,
  V.~A.~Kuzmin and I.~I.~Tkachev, Phys.\ Rev.\ \textbf{D36} (1987)
  2919. W.~Fischler, D.~Morgan and J.~Polchinski, Phys.\ Rev.\
  \textbf{D42} (1990) 4042 developed the quantization of false-vacuum
  bubbles using the Dirac formalism and the WKB approximation. The
  treatment of Kraus and Wilczek (Ref.\ \cite{K-W}) follows the methods
  of Fischler, Morgan and Polchinski. 
\bibitem{PH-coll} P.~H\'{a}j\'{\i}\v{c}ek, Commun.\ Math.\ Phys.\
  \textbf{150} (1993) 545.
\bibitem{K-W} P.~Kraus and F.~Wilczek, Nucl.\ Phys.\ \textbf{B433}
  (1995) 403.
\bibitem{M-Y} E.~A.~Martinez and J.~W.~York, Jr., Phys.\ Rev.\
  \textbf{D40} (1989) 2125.
\bibitem{berezin} V.~A.~Berezin, Phys.\ Lett.\ \textbf{B241} (1990)
  194.
\bibitem{ital} R.~Casadio and G.~Venturi, Class.\ Quantum Grav.\
  \textbf{13} (1996) 2715.
\bibitem{FGG} E.~Farhi, A.~H.~Guth and J.~Guven, Nucl.\ Phys.\
  \textbf{B339} (1990) 417. 
\bibitem{BGG} S.~K.Blau, E.~I.~Guendelman and A.~H.~Guth, Phys.\ Rev.\
  \textbf{D35} (1978) 1747.
\bibitem{H-K-K} P.~H\'{a}j\'{\i}\v{c}ek, B.~S.~Kay and K.~Kucha\v{r},
  Phys.\ Rev.\ \textbf{D46} (1992) 5439.
\bibitem{ADM} A.~Arnowitt, S.~Deser and C.~V.~Misner, in {\em
  Gravitation: an Introduction to Current Research}. Ed.\ by
  L.~Witten. Wiley, New York, 1962.
\bibitem{RT} T.~Regge and C.~Teitelboim, Ann.\ Phys.\ (N.Y.)
  \textbf{88} (1974) 286.
\bibitem{unique-H} D.~G.~Currie and E.~J.~Saletan, J. Math.\ Phys.\
  \textbf{7} (1966) 967; P.\ Havas, Acta Phys.\ Austr.\ \textbf{38}
  (1973) 145.
\bibitem{B-K} D.~Brown and K.~Kucha\v{r}, Phys.\ Rev.\ \textbf{D51}
  (1995) 5600.
\bibitem{dautcourt} R.~Dautcourt, Math.\ Nachr.\ \textbf{27} (1964) 277. 
\bibitem{israel} W.~Israel, Nuovo Cim.\ \textbf{44B} (1966) 1;
  \textbf{48B} (1967) 463.
\bibitem{D-tH} T.~Dray and G.~'t Hooft, Commun.\ Math.\ Phys.\ \textbf{99} 
  613 (1985); I.~H.~Redmount, Progr.\ Theor.\ Phys.\ \textbf{73}
  (1985) 1401.
\bibitem{B-I} C.~Barrab\`{e}s and W.~Israel, Phys.\ Rev.\ \textbf{D43}
  (1991) 1129. 
\bibitem{margu} P.~H\'{a}j\'{\i}\v{c}ek, Nucl.\ Phys.\ \textbf{B} (Proc.\
  Suppl.) \textbf{57} (1997) 115.

\end{thebibliography}
\end{document}